\documentclass[prd,superscriptaddress,nofootinbib]{revtex4}
\usepackage{amssymb,amsmath,graphicx,amscd,epsfig}
\usepackage[colorlinks=true, pdfstartview=FitV, linkcolor=blue, citecolor=red, urlcolor=magenta, breaklinks=true]{hyperref}

\newcommand{\be}{\begin{equation}}
\newcommand{\ee}{\end{equation}}
\newcommand{\beqq}{\setlength\arraycolsep{2pt}\begin{eqnarray}}
\newcommand{\eeqq}{\vspace{0cm} \end{eqnarray}}
\newcommand{\bea}{\begin{eqnarray}}
\newcommand{\eea}{\end{eqnarray}}

\begin{document}
\title{Stochastic motion in an expanding noncommutative fluid}

\author{M. A. Anacleto} \email{anacleto@df.ufcg.edu.br}
\affiliation{Departamento de F\'{\i}sica, Universidade Federal de Campina Grande 
Caixa Postal 10071, 58429-900 Campina Grande, Para\'iba, Brazil}

\author{C. H. G. Bessa } \email{chgbessa@yahoo.com}
\affiliation{Departamento de F\'{\i}sica, Universidade Federal de Campina Grande 
Caixa Postal 10071, 58429-900 Campina Grande, Para\'iba, Brazil}

\author{F. A. Brito}\email{fabrito@df.ufcg.edu.br}
\affiliation{Departamento de F\'{\i}sica, Universidade Federal de Campina Grande
Caixa Postal 10071, 58429-900 Campina Grande, Para\'iba, Brazil}
\affiliation{Departamento de F\'isica, Universidade Federal da Para\'iba, 
Caixa Postal 5008, 58051-970 Jo\~ao Pessoa, Para\'iba, Brazil}

\author{E. J. B. Ferreira}  \email{ewertonjeferson@hotmail.com}
\affiliation{Departamento de F\'{\i}sica, Universidade Federal de Campina Grande 
Caixa Postal 10071, 58429-900 Campina Grande, Para\'iba, Brazil}
\affiliation{Departamento de F\'isica, Universidade Federal da Para\'iba, 
Caixa Postal 5008, 58051-970 Jo\~ao Pessoa, Para\'iba, Brazil}

\author{E. Passos}\email{passos@df.ufcg.edu.br}
\affiliation{Departamento de F\'{\i}sica, Universidade Federal de Campina Grande
Caixa Postal 10071, 58429-900 Campina Grande, Para\'iba, Brazil}

%%%%%%%%%%%%%%%%%%%
\begin{abstract}
A model for an expanding noncommutative acoustic fluid analogous to a Friedmann-Robertson-Walker geometry is derived. For this purpose, a noncommutative Abelian Higgs model is considered in a (3+1)-dimensional spacetime. In this scenario, we analyze the motion of test particles in this fluid. The study considers a scalar test particle coupled to a quantized fluctuating massless scalar field. For all cases studied, we find corrections due to the noncommutativity in the mean squared velocity of the particles. The nonzero velocity dispersion for particles that are free to move on geodesics disagrees with the null result found previously in the literature for expanding commutative fluid. %We interpret this result as a piece of evidence for the noncommutativity.
\end{abstract}

\maketitle

\pretolerance10000

\section{Introduction}
Classical and quantum fluids have been considered in the literature as theoretical 
or experimental models where it is possible to mimic some effects present in theories of classical 
and quantum gravity, as well as, in quantum field theory in curved spacetime --- 
see Refs.~\cite{unruh81,Visser:1997ux,Barcelo2011,faccio2013,kms10,
fs09,wf20,doc12,bcl04,jwvg07,pfl10,blv11,blv03,ff03,Cha:2016esj,bffr08,foglizzo2012,tprw19,tprw20,pcrw18,pw20,ekjsc18,
lp19,sll19,Vieira:2014rva}. Some of these effects are related to Hawking radiation, particle production by cosmological expansion, superradiance, and so on. Moreover, experimental studies using classical and quantum fluids as table-top experiments have been done in recent years with remarkable advances. As examples, we mention the possible observation of classical superradiance~\cite{Torres2017} and Hawking radiation \cite{steinhauer2016,wtpul11,ngks19}. Another variety of analog model scenarios with remarkable experimental advances are also present in fiber-optics and slow light models \cite{lp00,l02,us03,pkrhkl08,su05,drbsl19}.  
In addition, based on the Abelian Higgs model, relativistic acoustic 
metrics~\cite{Bilic:1999sq,Ge:2010wx,Visser:2010xv}
were found in~\cite{Anacleto2012} for a noncommutative background, in~\cite{Anacleto:2010cr} for a Lorentz-violating background and in~\cite{Anacleto:2013esa} with terms of high derivatives in the bosonic sector.
Hence, studies related to Hawking radiation, entropy and superradiance were performed 
in~\cite{Anacleto:2011tr,Rinaldi:2011nb,Rinaldi:2011aa,Giovanazzi:2011az,Zhang:2011zzh,Anacleto:2012ba,Zhao:2012zz,Anacleto:2012du,Anacleto:2014apa,Anacleto:2015awa,
Anacleto:2016qll,Anacleto:2018acl,Anacleto:2019rfn}.

A distinct effect also mentionable is the stochastic motion of a particle under quantum fluctuations. This is a non-trivial quantum effects and it is shown that this motion can be induced by the presence of reflective plates \cite{gs99,jr92,YuFord2004,YuChen2004,Seriu2009,Seriu2008,hwl08,whl12,Lorenci2019,clrr19,pf11,lrs16,clrrs18}, non-trivial topology \cite{cy04,zy05,br20}, or even by a time-dependent  expanding universe \cite{Bessa2009,Bessa2017,mm19} without boundaries. In particular, in Ref. \cite{Bessa2009}, different types of classical particles\footnote{One example of the particles considered are point electric charges that could be under the influence of a classical external and non-fluctuating force, $f_{ext}$. They were named bound particles. Another example are particles that are free to follow their own geodesics with $f_{ext} = 0$, they were named free particles.} coupled to a fluctuating quantum electromagnetic field were considered in a spatially flat Friedmann-Robertson-Walker (FRW) universe. In this scenario, it was shown that the bound electric particle can undergo stochastic (Brownian) motion with a non-zero mean squared velocity (velocity dispersion) while the free electric particle had a null velocity dispersion. In a different scenario, see Ref.~\cite{Bessa2017}, it was proposed the possibility to observe this effect in an analog cosmological model by the use of a Bose-Einstein condensate (BEC) that simulates a spatially flat FRW geometry. To establish this analogy, a linearized perturbation in the field, which is present in the equation of motion of the BEC, was done and it was found that these perturbations describe the same equation of motion for a massless scalar field in curved spacetime \footnote{Recently, a model using BEC, which reproduces some behavior of an expanding universe was performed experimentally in Ref.~\cite{ekjsc18}.}. The main results calculated in that paper was that the velocity dispersion for a bound scalar particle (atoms that constitute the fluid) was non-null and for a free scalar particle a null dispersion was found, exactly to what was calculated in Ref. \cite{Bessa2009} for the charged particle. 

Another type of system considered in recent years is a noncommutative 
fluid~\cite{brito2016,Barosi:2008gx,Balachandran:2007ua,Chu:2000ww,Alexander:2001dr,Jackiw:2001dj,Das:2016hmc,Mitra:2018ezo,Das:2018puw}. In this case, the fluid description is valid only in energy scales much smaller than the Planck scale. In general the analogy is made with a Schwarzschild or Kerr-like geometry,  and the corrections to the Hawking temperature are in many cases derived \cite{Anacleto2012}. One of the purposes of this paper, Sec. \ref{sec_metric}, is to show that with the same noncommutative Lagrangian considered in Ref. \cite{Anacleto2012}, it is possible to acquire an analogous to a FRW geometry by the use of a linearized perturbation in the field. Thus, a natural question to ask is whether it is possible to observe the influence of the noncommutativity in the velocity dispersion  found in Ref. \cite{Bessa2017}. For this purpose, both the free and bound particles described above will be considered and no boundaries will be taken into account in the current study. Corrections in the velocity dispersion due to the noncommutative parameter ($\theta$) will be found for both particles. {The most important result is that the velocity dispersion, for the free particles, will be different from zero in this noncommutative fluid.  This reveals that free particles can possess stochastic motion induced by quantum fluctuations and that the noncommutativity of space plays a fundamental role in the model.   }

The rest of the paper is outlined as follows: in Sec. \ref{random FRW}, we calculate the formal expressions of the velocity dispersions in terms of integrals that depend on the scale factor of the analogous FRW geometry for both free and bound particles. The scale factor that describes the way the fluid expands is introduced in Sec. \ref{Sec.tanh} and, as proposed by many authors~\cite{jwvg07,pfl10,Bessa2017}, we consider an asymptotic expanded scale factor. Thus, the velocity dispersions for the free and bound particles are finally expressed. {The possible effects of metric fluctuations on the motion of the particles will be briefly discussed in Sec. \ref{Sec:fluctuations}.} In Sec. \ref{concl} the main results are summarized and some interpretations are mentioned. In this paper we use units where $\hbar = c = 1.$

\section{Acoustic metric for a noncommutative geometry}\label{sec_metric}

To obtain an analog Friedman-Robertson-Walker (FRW) geometry from a noncommutative fluid, we start with the Lagrangian for a noncommutative Abelian Higgs model in flat spacetime  modified in the scalar and gauge sector \cite{Anacleto2012,Ghosh:2004wi,sw99} 
\begin{eqnarray}\label{Moyal}
{ \mathcal{\hat{L}} = -\dfrac{1}{4}\hat{F}_{\mu\nu}\ast\hat{F}^{\mu\nu} + (D_\mu\hat{\phi})^{\dagger}\ast D^\mu\hat{\phi} + m^2{\hat{\phi}}^{\dagger}\ast\hat{\phi}  - b\hat{\phi}^{\dagger}\ast \hat{\phi}\ast\hat{\phi}^{\dagger}\ast \hat{\phi}   },
\end{eqnarray}
{where the Moyal product was used.   In what follows, we apply a Seiberg-Witten map \cite{sw99}}
\begin{align}\label{Witten}
\nonumber & { {\hat{A}}_{\mu} = A_{\mu} + \theta^{\nu\rho}A_{\rho}(\partial_\nu A_\mu - \frac{1}{2}\partial_{\mu}A_\nu)  }, \\ 
          & { {\hat{F}}_{\mu\nu} = F_{\mu\nu} + \theta^{\rho\alpha}(F_{\mu\rho}F_{\nu\alpha} +  A_\rho + \partial_{\alpha}F_{\mu\nu})   }, \\
\nonumber & { {\hat{\phi}} = \phi - \frac{1}{2}\theta^{\rho\alpha}A_\rho\partial_{\alpha}\phi    }, 
\end{align}
{where, only the lowest order terms in $\theta^{\mu\nu}$ are taken into account. From Eqs. (\ref{Moyal}) and (\ref{Witten}) we see that the noncommutativity is coupled via the electromagnetic field. So we obtain the following Lagrangian} 
\begin{eqnarray}\label{eq00}
\mathcal{\hat{L}}=&-&\dfrac{1}{4}F_{\mu\nu}F^{\mu\nu}\left(1+\dfrac{1}{2}\theta^{\alpha\beta}F_{\alpha\beta}\right) + \left(1-\dfrac{1}{4}\theta^{\alpha\beta}F_{\alpha\beta}\right)(|D_{\mu}\phi|^{2}+m^{2}|\phi|^{2}-b|\phi|^{4})\nonumber\\
&+&\dfrac{1}{2}\theta^{\alpha\beta}F_{\alpha\mu}\left[(D_{\beta}\phi)^{\dagger}D^{\mu}\phi+(D^{\mu}\phi)^{\dagger}
D_{\beta}\phi\right],
\end{eqnarray}
where the operator $D_{\mu}=\partial_{\mu}-ieA_{\mu}$ and $F_{\mu\nu}=\partial_{\mu}A_{\nu}-\partial_{\nu}A_{\mu}$. Here, $F_{\mu\nu}$ is the Maxwell tensor, $A_\mu$ is the {4-potential, and $e$, $b$ are} coupling constants. The term $\theta^{\alpha\beta}$ is the real constant noncommutative parameter with dimensions of length squared represented by an anti-symmetric $D$-dimensional square matrix~\cite{Szabo:2001kg,Nicolini:2008aj}. The field $\phi$ can be decomposed by  $\phi = \sqrt{\rho(\bar{x},\bar{t})}e^{iS(\bar{x},\bar{t})}$, where $\rho(\bar{x}, \bar{t})$ is the fluid density and $S(\bar{x}, \bar{t})$   is a phase. We consider that the noncommutative effect is absent in the time coordinate.  Following Ref. \cite{Anacleto2012} let us apply the  perturbations in {$\rho=\rho_{0}+\rho_{1}$, $S=S_{0}+S_{1}$ and $\phi = \phi_0 + \phi_1$ in Eq. (\ref{eq00}), where $\rho_1\ll\rho_0, S_1\ll S_0$ and consequently $\phi_1\ll\phi_0$ . When we compute the equations of motion with these perturbations, it can be viewed as a Klein-Gordon equation in curved spacetime and the following noncommutative relativistic metric can be given by Eq. (31) of Ref. \cite{Anacleto2012} :}
\begin{eqnarray}\label{eqmetric1}
d{\bar{s}}^{2}=\dfrac{b\rho_{0}}{2c_{s}\sqrt{f}}\left[ - \mathcal{F}(v)d{t'}^{2}+ \Lambda\left(\dfrac{v^{i}v^{j}\Gamma+\Sigma^{ij}}{\Lambda\mathcal{F}(v)} + \delta^{ij}\right)d{\bar{x}}^{i}d{\bar{x}}^{j}\right],
\end{eqnarray}
with the following terms
\begin{eqnarray}
dt' &=& d\bar{t}+\dfrac{\vec{\xi}(v)\cdot d\vec{\bar{x}}}{2\mathcal{F}(v)}, \nonumber\\
f&=& [(1-2\vec{\theta}\cdot\vec{B})(1+c_{s}^{2})-(1+4\vec{\theta}\cdot\vec{B})v^{2}] -3(\vec{\theta}\times\vec{E})\cdot\vec{v} +2(\vec{B}\cdot\vec{v})(\vec{\theta}\cdot\vec{v}),\nonumber \\
\mathcal{F}(v)&=& (1-3\vec{\theta}\cdot\vec{B})c_{s}^{2}-(1+3\vec{\theta}\cdot\vec{B})v^{2}
-(\vec{\theta}\times\vec{E})\cdot\vec{v} +2(\vec{\theta}\cdot\vec{v})(\vec{B}\cdot\vec{v}),\nonumber\\
\Lambda(v)&=& (1+\vec{\theta}\cdot\vec{B})(1+c_{s}^{2}-v^{2})-(\vec{\theta}\times\vec{E})\cdot\vec{v},\\
\vec{\xi}(v)&=& [2(1+2\vec{\theta}\cdot\vec{B})-(\vec{\theta}\times\vec{E})\cdot\vec{v}]\vec{v}
+(1+c_{s}^{2})(\vec{\theta}\times\vec{E})-(\vec{B}\cdot\vec{v})\vec{\theta}-(\vec{\theta}\cdot\vec{v})\vec{B},\nonumber\\
\Gamma(v)&=& 1+4\vec{\theta}\cdot\vec{B}+(1+2\vec{\theta}\cdot\vec{B})c_{s}^{2}-(1+4\vec{\theta}\cdot\vec{B})v^{2}-2(\vec{\theta}\times\vec{E})\cdot\vec{v}+2(\vec{\theta}\cdot\vec{v})(\vec{B}\cdot\vec{v}),\nonumber\\
\Sigma^{ij}(v)&=&[(1+c_{s}^{2})(\vec{\theta}\times\vec{E})^{i}-(\vec{B}\cdot\vec{v})\theta^{i}-(\vec{\theta}\cdot\vec{v})B^{i}]v^{j}\nonumber,
\end{eqnarray}
%where  $c_{s}$  is the local sound velocity in the fluid and $\vec{v}$ is the velocity flux with $\vec{E}$ and $\vec{B}$ beeing the electric and magnetic fields, respectively. 
{where  $c_{s}^2 = b\rho_{0}/2\mathcal{W}_{0}^{2}$ is the local sound velocity in the fluid and $\vec{v}=\vec{v}_{0}/\mathcal{W}_{0}$ is the velocity flux, with $\mathcal{W}_{0} = -\dot{S} + eA_{t}$ and $\vec{v}_{0} = \nabla S_{0}+e\vec{A}$ (the local velocity field) and $b$ is a coupling constant with $\vec{E}$ and $\vec{B}$ being the electric and magnetic fields, respectively.}

{To simplify our expressions, let us choose a null electric field ($\vec{E} = 0$). Now the noncommutativity is coupled only to the magnetic field, and considering the constant coupling $e=0$  with a phase $S_0$ being time-dependent only, $S_0=S_0(t)$, the velocity flux ($\vec{v}$) is now equal to zero. Thus, considering that $c_{s}^{2}\ll 1$,  we obtain the non-relativistic acoustic metric}
\begin{equation}
d\overline{s}^{2}=\dfrac{b\rho_{0}}{2}\left[-\dfrac{(1-3\vec{\theta}\cdot\vec{B})}{(1-2\vec{\theta}\cdot\vec{B})^{\frac{1}{2}}}
{  {   c_{s}^2  }}     d\overline{t}^{ 2}+\dfrac{(1+\vec{\theta}\cdot\vec{B})}{(1-2\vec{\theta}\cdot\vec{B})^{\frac{1}{2}}}c_{s}^{-1}\delta^{ij}d\overline{x}^{i}d\overline{x}^{j}\right].
\end{equation}
By defining the following change in the spatial coordinates $ d\overline{x}^{i}=\left(\frac{H}{W}\right)dx^{i}$, we obtain  
\begin{equation}\label{eqmetric2}
d\overline{s}^2=\dfrac{b\rho_0}{2} H^{2}\left[    {{    -c_{s}^2      }}  d\overline{t}^2 + c_{s}^{-1}\delta^{ij}dx^{i}dx^{j}\right],
\end{equation}
where $H$ and $W$ were defined by
\begin{eqnarray}\label{eq:HW}
H^{2}&=&\dfrac{(1-3 \vec{\theta}\cdot\vec{B})}{(1-2\vec{\theta}\cdot\vec{B})^{\frac{1}{2}}},\nonumber \\
W^{2}&=&\dfrac{(1+ \vec{\theta}\cdot\vec{B})}{(1-2\vec{\theta}\cdot\vec{B})^{\frac{1}{2}}}.
\end{eqnarray}
The upper bar in the previous expressions has been added to distinguish the variables in the mathematical manipulations. Note that, in this model, the magnetic field ${\vec{B}}$ plays the role to turn on and off the noncommutativity. If the noncommutativity is off or null, the terms $H$ and $W$ are equal to the unity. In fact the term $\theta$ must be very small which in turn makes $H \approx W \approx 1$.

In our model, we take the sound velocity to be time dependent, $c_{s}=c_{s}(t)$,  and it can be represented in terms of the parameter $\chi(t)=\left[\frac{c_{s}(t)}{c_0}\right]^{2}$, which is interpreted as a scale factor of the expanding fluid. Note that, one can admit, without loss of generality,  that for a given initial time $t_0$, $\chi(t_0) = 1$. The line element given by Eq. (\ref{eqmetric2}) becomes  
\begin{eqnarray}\label{eqmetric3}
d\overline{s}^2=\dfrac{b\rho_0}{2} H^{2}\left[-\chi^{\frac{1}{2}}(t)  {{     c_{0}^2      }}     d\overline{t}^2 + \chi^{-\frac{1}{2}}(t)c_{0}^{-1}\delta^{ij}dx^{i}dx^{j}\right],
\end{eqnarray}
that defining $\Omega_{0}^{2}=\frac{b\rho_0}{2c_0}$ and $ds^{2} = \Omega_{0}^{-2}d\overline{s}^{2}$, by performing a change in the time coordinate, such that $dt=\chi^{\frac{1}{4}}(t)d\overline{t}$, we have
\begin{eqnarray}\label{eq:01} %MÉTRICA NO TEMPO COORDENADO%
ds^{2}=H^{2}\left[-c_{0}^{2}dt^2 + a^{2}(t)\delta^{ij}dx^{i}dx^{j}\right],
\end{eqnarray}
where
\begin{eqnarray}\label{eq:14}
a^{2}(t)=\chi^{-\frac{1}{2}}(t) = \dfrac{c_{0}}{c_{s}(t)}.
\end{eqnarray}
Apart from the constant $H$, these definitions gives an effective metric that mimics a FRW geometry for a certain coordinate time $t$. Since $a(\eta) = c_{0}/c_{s}(t)$, it is important to note that { $c_{s}(t=t_f)<c_{0}$, where $t_f$ is the final time when the expansion ends,} is a necessary condition to the expansion occur, that is, for $a(t)$ assuming increasing values. In order to get a conformal effective metric, one  defines the following transformation in the time coordinate $dt = ad\eta$, where $\eta$ is the conformal time and the effective conformal metric is
\begin{eqnarray}\label{eq:02} %MÉTRICA NO TEMPO CONFORME%
ds^{2}=H^{2}a^{2}(\eta)\left[-c_{0}^{2}d\eta^{2} + \delta^{ij}dx^{i}dx^{j}\right],
\end{eqnarray}
with $a^{2}(\eta)H^2$ being the conformal factor. When $H$ is equal to the unity,  we recover the usual forms of Eqs. (\ref{eq:01}) and (\ref{eq:02}). This form of the metric is conformal to Minkowski spacetime and it will be useful when we evaluate the two point function for a scalar field  in the next section.

\section{Random motion of particles in an analog FRW noncommutative geometry}\label{random FRW}

The motion of a point scalar particle with mass $m$ in a conformal curved space-time is represented by the following equation %\cite{ppv11}

\begin{eqnarray}\label{eqfcommu}
{f}^\mu = m\dfrac{Du^{\mu}}{d\tau} = qg^{\mu\nu}\nabla_{\nu}\phi,
\end{eqnarray}
where $f^\mu$ is the 4-force, $u^\mu$ is the 4-velocity of the scalar particles, 
$ q $ is the charge of the scalar particle interacting
with a massless scalar field and the operator $D/d\tau$ is the covariant derivative. This equation is valid for a commutative space. However, in a noncommutative space it must be modified. 

{For the case studied here, we consider the motion of free and bound scalar test particles described before. They are the constituents of a noncommutative fluid that expands according to the metric given by Eq. (\ref{eq:02}). Thus, these point-like massive particles interact with a massless scalar field of the Abelian-Higgs model performed by the Lagrangian (\ref{eq00}). The acoustic perturbations of this system (i.e., phonons) are described by this field,  and the equation of motion of a single particle with mass $m$ in such conformal curved space-time (\ref{eq:02}) is given by  }
\begin{eqnarray}\label{eq:03}
f^{\mu}=q\left[ \left(1-\dfrac{1}{4}\theta^{\alpha\beta}F_{\alpha\beta}\right)g^{\mu\nu}+\Theta^{\mu\nu}\right]\nabla_{\nu}\phi,
\end{eqnarray}
{where the Lagrangian given by Eq. \eqref{eq00} was used. This is the equation of motion to the scalar particle in the noncommutative space with the metric $g^{\mu\nu}$ given by Eq. \eqref{eq:01} or \eqref{eq:02} and }

\begin{eqnarray}
\Theta^{\mu\nu} = \theta^{\alpha\mu}F_{\alpha}^{\phantom{a}\nu}.
\end{eqnarray}
Note that when $\theta = 0$ we recover Eq. (\ref{eqfcommu}).

Adopting a particular $i$-direction and considering only non-commutative effects on spatial part, i.e., $\theta^{0j} = \theta^{i0} = 0$, Eq. \eqref{eq:03} becomes
\begin{equation}\label{eq:04}
f^{i}=q\left[\left(1-\dfrac{1}{2}\vec{\theta}\cdot\vec{B} \right)g^{ip}+\Theta^{ip}\right]\nabla_{p}\phi,
\end{equation}
where $i,p=x, y, z,$
\begin{equation}\label{eq:05}
\Theta^{ip}=\theta^{ji}F_{j}^{\ p},
\end{equation}
and
\begin{eqnarray}\label{eq:06}
\theta^{ij} = \epsilon^{ijk}\theta^{k},& \ & F^{ij}= \epsilon^{ijl}B^{l},
\end{eqnarray} 
where a sum is adopted in repeated indices. Thus, if $B^i=0$ or $\theta=0$, Eq. (\ref{eqfcommu}) is recovered.%Note that when $\theta = 0$,  Eq. (\ref{eqfcommu}) is recovered.

Following the same procedure of Refs. \cite{Bessa2017} and \cite{Bessa2009}, with the effective metric given by  Eq. (\ref{eq:01}), we obtain 
\begin{eqnarray}
\dfrac{1}{m}f^{i}=\frac{Du^\mu}{dt}=\dfrac{du^{i}}{dt}+2\dfrac{\dot{a}}{a}u^{i},
\end{eqnarray}
where $\dot{a}=da/dt$ and we have assumed non-relativistic motion for the particles, which  implies that the time coordinate $t$ is their proper time $\tau$.

In the sequence, we will consider that the force $f^{i}$  can be split into two parts, the first one is originated by a non-fluctuating classical external force ($f^{i}_{ext}$) and the second one is a fluctuating force ($f^{i}_{q}$) associated with the quantized scalar field. Thus, we have
\begin{equation}\label{eq:fcfq}
\dfrac{1}{m}\left(f^{i}_{ext}+f^{i}_{q}\right)=\dfrac{du^{i}}{dt}+2\dfrac{\dot{a}}{a}u^{i}.
\end{equation}

In the equation above we can study two distinct situations. In the first case, we consider free particles which are characterized by a null external force ($ f^{i}_{ext} = 0$). Thus, they can move freely following their geodesic in the expanding background. In the second case, they are named `bound particles'. Now, they are influenced by an external force given by $f^{i}_{ext} = 2m\dfrac{\dot{a}}{a}u^{i}$. This force cancels out locally the effects of the expansion. 
Both cases will be study in the next two sections. We will find that the non-commutativity can give relevant contribution to the stochastic motion of the particles in this expanding background.

%%%%%%%%%%%%%%%%%%%%%%%%%%%%%%%%%%%%%%%%%%%%%%%%%%%%%%%%%%%%%%%%%%%%%%%%%%%%%%%%%%%%%%%%%%%%%%%%%%%%%%%%%%%%

\subsection{Free particle}
In this section we consider that no external, classical force, is  acting on the particle (i.e., $f^{i}_{ext} = 0$). We implement this into Eq. \eqref{eq:fcfq} to obtain the following equation\footnote{In what follows, the sub-indexes $q$ from $f_q$ will be omitted.    } 
\begin{eqnarray}
\dfrac{1}{m}f^{i}=\dfrac{1}{a^{2}}\dfrac{d}{dt}(a^{2}u^{i}),
\end{eqnarray}
that by integrating once and admitting that the particle is initially at rest ($u^i(t_{0}) = 0$), we have 
\begin{eqnarray}
u^{i}(t_f,r)&=&\dfrac{1}{ma^{2}(t_f)}\int^{t_f}_{t_0}a^{2}(t)f^{i}(t,r)dt.
\end{eqnarray}
From the above results, the correlation function for the velocity of the particle is
\begin{eqnarray}\label{eq:07}
\langle (\Delta u^i)^{2}\rangle =\dfrac{1}{m^{2}a^{4}(t_f)}\iint dt_{1}dt_{2}a^{2}(t_1)a^{2}(t_2)\langle f^{i}(t_1,r_1)f^{i}(t_2,r_2) \rangle_{_{FRW}}.
\end{eqnarray}

Now we follow the standard procedure \cite{YuFord2004} which consists in assuming that the field $\phi$ can be decomposed into a classical and a quantum part, i.e., $\phi = \phi_{c} + \phi_q$ and the conditions $\langle \phi(t_1,r_1)\rangle=0$, $ \langle \phi(t_1,r_1)\phi(t_2,r_2) \rangle \neq 0$ are satisfied.  In this way, one can use the fact that $f^i$ is related to a scalar field by Eq.~(\ref{eq:04}). Consequently, the equation above becomes
\begin{eqnarray}\label{eq:08}
\langle (\Delta u^i)^{2}\rangle &=&\dfrac{q^{2}H^{-4}}{m^{2}a^{4}(t_f)}\left(1-\dfrac{1}{2}\vec{\theta}\cdot\vec{B}\right)^{2}\iint dt_{1}dt_{2}\partial_{i_1}\partial_{i_2} {{     \langle\phi(t_1, r_1)\phi(t_2, r_2)\rangle_{_{FRW}}          }}\nonumber \\
&+&\dfrac{q^{2}H^{-2}\Theta^{is}}{m^{2}a^{4}(t_f)}\left(1-\dfrac{1}{2}\vec{\theta}\cdot\vec{B}\right) \iint dt_{1}dt_{2}a^{2}(t_2)\partial_{i_1}\partial_{s_2}               {{     \langle\phi(t_1, r_1)\phi(t_2, r_2)\rangle_{_{FRW}}          }}\nonumber \\
&+&\dfrac{q^{2}H^{-2}\Theta^{ip}}{m^{2}a^{4}(t_f)}\left(1-\dfrac{1}{2}\vec{\theta}\cdot\vec{B}\right) \iint dt_{1}dt_{2}a^{2}(t_1)\partial_{p_1}\partial_{i_2}               {{     \langle\phi(t_1, r_1)\phi(t_2, r_2)\rangle_{_{FRW}}          }} \nonumber \\
&+&\dfrac{q^{2}\Theta^{ip}\Theta^{is}}{m^{2}a^{4}(t_f)}\iint dt_{1}dt_{2}a^{2}(t_1)a^{2}(t_2)\partial_{p_1}\partial_{s_2}
               {{     \langle\phi(t_1, r_1)\phi(t_2, r_2)\rangle_{_{FRW}}          }},
\end{eqnarray}
where  we have assumed that $H$ is a constant as well as $\vec{\theta}$ and $\vec{B}$, {and the subscript $FRW$ in $\langle \phi(\eta_1, r_1)\phi(\eta_2, r_2)\rangle_{FRW} $ indicates that the vacuum expectation value in the FRW-geometry was taken}. In this case, there are no  distinction between these parameters at different times, e.g., $H_{1}=H_{2}$, ${\theta}_{1} = {\theta}_{2}$, and ${B}_{1}={B}_{2}$.
Note that to obtain the proper velocity from Eq. \eqref{eq:08} we use the relation between the coordinate $x^{i}$ and the proper distance $l^{i}$, in which  $l^{i}=a(t_{f})x^{i}$, where $a(t_{f})$ is the scale factor in certain final time. Thus, the proper velocity  $v^{i}$ is related to $u^{i}$ by $u^{i}a(t_ {f}) = v^{i}$. 
In this way we obtain
\begin{equation}\label{eq:09}
\langle (\Delta u^i)^{2}\rangle = \frac{1}{a^{2}(t_f)H^{2}}\langle (\Delta v^i)^{2}\rangle,
\end{equation}
and the proper velocity dispersion of the particle is
\begin{eqnarray}\label{eq:10}
\langle (\Delta v^i)^{2}\rangle &=&\dfrac{q^{2}H^{-2}}{m^{2}a^{2}(t_f)}\left(1-\dfrac{1}{2}\vec{\theta}\cdot\vec{B}\right)^{2}\iint dt_{1}dt_{2}\partial_{i_1}\partial_{i_2} {{     \langle\phi(t_1, r_1)\phi(t_2, r_2)\rangle_{_{FRW}}        }}\nonumber \\
&+&\dfrac{q^{2}\Theta^{is}}{m^{2}a^{2}(t_f)}\left(1-\dfrac{1}{2}\vec{\theta}\cdot\vec{B}\right) \iint dt_{1}dt_{2}a^{2}(t_2)\partial_{i_1}\partial_{s_2}               {{     \langle\phi(t_1, r_1)\phi(t_2, r_2)\rangle_{_{FRW}}        }}\nonumber \\
&+&\dfrac{q^{2}\Theta^{ip}}{m^{2}a^{2}(t_f)}\left(1-\dfrac{1}{2}\vec{\theta}\cdot\vec{B}\right) \iint dt_{1}dt_{2}a^{2}(t_1)\partial_{p_1}\partial_{i_2}               {{     \langle\phi(t_1, r_1)\phi(t_2, r_2)\rangle_{_{FRW}}        }}\nonumber \\
&+&\dfrac{q^{2}H^{2}\Theta^{ip}\Theta^{is}}{m^{2}a^{2}(t_f)}\iint dt_{1}dt_{2}a^{2}(t_1)a^{2}(t_2)\partial_{p_1}\partial_{s_2}
               {{     \langle\phi(t_1, r_1)\phi(t_2, r_2)\rangle_{_{FRW}}        }}. 
\end{eqnarray}
Using the relation present in Ref. \cite{Birrell1984}, we obtain
\begin{equation}\label{eq:11}
\langle\phi(\eta_{1},r_{1})\phi(\eta_{2},r_{2})\rangle_{_{FRW}}=H^{-2}a^{-1}(\eta_{1})a^{-1}(\eta_{2})\langle\phi_{1}(\eta_{1},r_{1})\phi_{2}(\eta_{2},r_{2})\rangle_{_M},
\end{equation} 
which relates the  two-point (Hadamard) function of a massless scalar field  in the conformal FRW spacetime to the two-point function in Minkowski spacetime.

Substituting \eqref{eq:11} into \eqref{eq:10} in terms of the conformal time ($dt=a(\eta)d\eta$), we obtain
\begin{eqnarray}\label{eq:12}
\langle (\Delta v^i)^{2}\rangle &=&\dfrac{q^{2}H^{-4}}{m^{2}a_{f}^{2}}\left(1-\dfrac{1}{2}\vec{\theta}\cdot\vec{B}\right)^{2}\iint d\eta_{1}d\eta_{2}\partial_{i_1}\partial_{i_2}           {{     \langle\phi(\eta_1, r_1)\phi(\eta_2, r_2)\rangle_{_{M}}     }}\nonumber \\
&+&\dfrac{q^{2}H^{-2}\Theta^{is}}{m^{2}a_{f}^{2}}\left(1-\dfrac{1}{2}\vec{\theta}\cdot\vec{B}\right) \iint d\eta_{1}d\eta_{2}a^{2}(\eta_{2})\partial_{i_1}\partial_{s_2}               {{     \langle\phi(\eta_1, r_1)\phi(\eta_2, r_2)\rangle_{_{M}}     }}\nonumber \\
&+&\dfrac{q^{2}H^{-2}\Theta^{ip}}{m^{2}a_{f}^{2}}\left(1-\dfrac{1}{2}\vec{\theta}\cdot\vec{B}\right) \iint d\eta_{1}d\eta_{2}a^{2}(\eta_{1})\partial_{p_1}\partial_{i_2}               {{     \langle\phi(\eta_1, r_1)\phi(\eta_2, r_2)\rangle_{_{M}}     }}\nonumber \\
&+&\dfrac{q^{2}\Theta^{ip}\Theta^{is}}{m^{2}a_{f}^{2}}\iint d\eta_{1}d\eta_{2}a^{2}(\eta_{1})a^{2}(\eta_{2})\partial_{p_1}\partial_{s_2}
                             {{     \langle\phi(\eta_1, r_1)\phi(\eta_2, r_2)\rangle_{_{M}}     }},
\end{eqnarray}
where $a(t_{f}) \equiv a_{f}$ is the scale factor in a final time when the expansion ends and the subscript $M$ indicates that the vacuum state is the Minkowski vacuum state. 
However, the equivalent Hadamard function for a massless scalar field is given by
\begin{eqnarray}\label{eq:f2p}
\langle\phi_{1}(\eta_{1},r_{1})\phi_{2}(\eta_{2},r_{2})\rangle_{M} = \dfrac{1}{4\pi^{2}}\left[ \dfrac{1}{-c_{0}^{2}(\eta_{1}-\eta_{2})^{2}+r^{2}} \right].
\end{eqnarray} 
Here $c_{0}$ is the speed of sound when the expansion starts and because no boundary is present, the  spatial separation $r$ is given by
\begin{eqnarray}\label{eq:fp1}
r^{2}&=& \Delta x^{2}+\Delta y^{2}+\Delta z^{2}=(x_{1}-x_{2})^{2}+(y_{1}-y_{2})^{2}+(z_{1}-z_{2})^{2}.
\end{eqnarray}
In Sec. \ref{Sec.tanh} we will evaluate the velocity dispersion  given by Eq. (\ref{eq:12}) for a noncommutative fluid that expands asymptotically.    

\subsection{Bound particles}

In this section let us admit that the particle is subject to a classical external non-fluctuating force of the type 
\begin{equation}
f^{i}_{ext}=2m\dfrac{\dot{a}}{a}u^{i}.
\end{equation} 
By substituting this into Eq. \eqref{eq:fcfq} and taking an integral, we obtain 
\begin{equation}
u^{i}(t_f,r)= \dfrac{1}{m}\int f^{i}(t,r)dt,
\end{equation}
where we considered a null velocity for the initial time $t_{0}$. Thus, the velocity dispersion of the bound particles is
\begin{eqnarray}
\langle(\Delta u^i)^{2}\rangle=\dfrac{1}{m^2}\iint dt_{1}dt_{2}\langle f^{i}(t_1,r_1)f^{i}(t_2,r_2)\rangle_{_{FRW}}.
\end{eqnarray}
Now using Eq.~\eqref{eq:04} and the metric (\ref{eq:01}),  we obtain
\begin{eqnarray}
\langle (\Delta u^i)^{2}\rangle &=&\dfrac{q^{2}H^{-4}}{m^{2}}\left(1-\dfrac{1}{2}\vec{\theta}\cdot\vec{B}\right)^{2}\iint dt_{1}dt_{2}a^{-2}(t_1)a^{-2}(t_2)\partial_{i_1}\partial_{i_2}         {{     \langle\phi(t_1, r_1)\phi(t_2, r_2)\rangle_{_{FRW}}       }}\nonumber \\
&+&\dfrac{q^{2}H^{-2}\Theta^{is}}{m^{2}}\left(1-\dfrac{1}{2}\vec{\theta}\cdot\vec{B}\right)\iint dt_{1}dt_{2}a^{-2}(t_1)\partial_{i_1}\partial_{s_2}
                                     {{     \langle\phi(t_1, r_1)\phi(t_2, r_2)\rangle_{_{FRW}}       }}\nonumber \\
&+&\dfrac{q^{2}H^{-2}\Theta^{ip}}{m^{2}}\left(1-\dfrac{1}{2}\vec{\theta}\cdot\vec{B}\right) \iint dt_{1}dt_{2}a^{-2} (t_2)\partial_{p_1}\partial_{i_2}                                    {{     \langle\phi(t_1, r_1)\phi(t_2, r_2)\rangle_{_{FRW}}        }}\nonumber \\
&+&\dfrac{q^{2}\Theta^{ip}\Theta^{is}}{m^{2}}\iint dt_{1}dt_{2}\partial_{p_1}\partial_{s_2}
                                     {{     \langle\phi(t_1, r_1)\phi(t_2, r_2)\rangle_{_{FRW}}       }}.
\end{eqnarray}
Following the same steps of the previous section we apply a conformal transformation  in time and using Eqs. \eqref{eq:09} and \eqref{eq:11} we get
\begin{eqnarray}\label{eq:23}
\langle (\Delta v^i)^{2}\rangle &=&\dfrac{q^{2}a^{2}_{f}H^{-4}}{m^{2}}\left(1-\dfrac{1}{2}\vec{\theta}\cdot\vec{B}\right)^{2}\iint d\eta_{1}d\eta_{2}a^{-2}(\eta_1)a^{-2}(\eta_2)\partial_{i_1}\partial_{i_2}        {{     \langle\phi(\eta_1, r_1)\phi(\eta_2, r_2)\rangle_{_{M}}      }}\nonumber \\
&+&\dfrac{q^{2}a^{2}_{f}H^{-2}\Theta^{is}}{m^{2}}\left(1-\dfrac{1}{2}\vec{\theta}\cdot\vec{B}\right)\iint d\eta_{1}d\eta_{2}a^{-2}(\eta_1)\partial_{i_1}\partial_{s_2}                                           {{     \langle\phi(\eta_1, r_1)\phi(\eta_2, r_2)\rangle_{_M}        }} \nonumber \\
&+&\dfrac{q^{2}a^{2}_{f}H^{-2}\Theta^{ip}}{m^{2}}\left(1-\dfrac{1}{2}\vec{\theta}\cdot\vec{B}\right) \iint d\eta_{1}d\eta_{2}a^{-2} (\eta_2)\partial_{p_1}\partial_{i_2}                                           {{     \langle\phi(\eta_1, r_1)\phi(\eta_2, r_2)\rangle_{_M}        }}\nonumber \\
&+&\dfrac{q^{2}a^{2}_{f}\Theta^{ip}\Theta^{is}}{m^{2}}\iint d\eta_{1}d\eta_{2}\partial_{p_1}\partial_{s_2}
                                                          {{     \langle\phi(\eta_1, r_1)\phi(\eta_2, r_2)\rangle_{_M}        }}.
\end{eqnarray}

{ In the next section, we will evaluate the velocity dispersion for free and bound particles for a fluid that expands asymptotically. However, it is worth noting that, when we quantize the scalar field the fluctuations related to this field could imply fluctuations in the effective metric (\ref{eq:01}).These metric fluctuations are viewed in general as linearized perturbations ($\gamma_{\mu\nu}$) upon the effective metric, see for instance Refs. \cite{f95,fs97}. So in this context, Eq. (\ref{eq:01}) should read,}
\begin{equation}\label{Eq:perturbation}
ds^2 =H^2(g_{\mu\nu}dx^{\mu}dx^{\nu} + \gamma_{\mu\nu}dx^{\mu}dx^{\nu}) 
\end{equation}
{where $g_{\mu\nu} = diag[-c_0, a^2, a^2, a^2], x^\mu \in \{t, x, y , z\}$ and the entries of the matrix $\gamma_{\mu\nu}$ are $0 < |\gamma_{\mu\nu}| \ll 1$. A complete treatment to our case is not an easy task once it involves the integration of Eq. (20). In section \ref{Sec:fluctuations}, we will give a simple example where the metric fluctuations are considered in the context of the analog models program, in which these fluctuations can induce sound cone fluctuations.We will see that such fluctuations should contribute as a minor or null deviation in the random motion of the particles. }

\section{Asymptotic expansion for a noncommutative fluid}\label{Sec.tanh}

In this section we investigate the behavior of the particle under the scenarios discussed in the previous section. We consider that the noncommutative fluid is taking an asymptotic expansion according to the following scale factor:  
\begin{equation}\label{eq:fatesc}
a^{2}(\eta)=a_{0}^{2}+a_{1}^{2}\tanh\left(\dfrac{\eta}{\eta_0}\right),
\end{equation}
which describes an asymptotically flat spacetime in extreme regions, where the  constant $a_{0}$ produces a vertical displacement of the point on the $a^{2}(\eta)$ axis, and $a_{1}$ modifies the spacing between asymptotic limits (maximum and minimum points of the $a^{2}(\eta)$). The parameter $\eta_{0}$ modifies the smoothness of the transition between the asymptotic regions but without changing the spacing between the maximum and minimum points.%($\eta^{2}_{0}+\eta^{2}_{1}$) and minimum ($\eta^{2}_{0}-\eta^{2}_{1}$) points.

Based on the asymptotic behavior of the scale factor we can write
\begin{eqnarray}\label{eq:131}
a_{0}^{2}=\dfrac{a_{f}^{2}+a_{i}^{2}}{2},
\end{eqnarray}
and 
\begin{eqnarray}\label{eq:132}
a_{1}^{2}=\dfrac{a_{f}^{2}-a_{i}^{2}}{2},
\end{eqnarray}
where $a_{i}$ and $a_{f}$ are the scale factor at the beginning and end of the expansion.
Note that, for $\chi(t=t_0)=1$ in Eq.~\eqref{eq:14}, we obtain $a_{i}=a(\eta=\eta_i)=1$.
Next, we will use this scale factor to evaluate the dispersion velocity for the free and bound particles, respectively.

\subsection{Free particle in an expanding noncommutative fluid}\label{Sec.tanh.free}

According to our choices in Eqs.~\eqref{eq:HW}, \eqref{eq:05}, and \eqref{eq:06}, the magnetic field is responsible to turn on the noncommutativity. To simplify our expressions, let us consider that it is turn on  in just one direction. Let us choose, for instance, $\vec{B} = B^z\hat{k}$. So, the dispersion in the $z$-direction is now parallel to the field vector. 
%Let us consider the dispersion in the $z$ direction, which is the direction parallel to the magnetic field ($\vec{B}$). According to Eqs.~\eqref{eq:HW}, \eqref{eq:05}, and \eqref{eq:06}, this is the field that turns on the noncommutativity. % in Eq. \eqref{eq:12}. 
Thus, when the field is off,   all terms proportional to $\theta$ disappear in Eq.~\eqref{eq:12} and only the first integral in the right hand side remains. This recover the result found in Ref.~\cite{Bessa2017}. However, as the scale factor does not appeas in the integrand, this integral gives infinite contribution and must be renormalized.  This procedure consists in subtracting the Minkowski contribution from the Minkowski two-point function. Thus,  the first integral gives null contribution.%Thus,  such integral this gives a trivial or null contribution. This is the same result found in Ref.~\cite{Bessa2017} for a commutative fluid. 

Now when the field $\vec{B}$ is on, the terms proportional to $\theta$ become important and we have to consider all terms with at least one $a^2(\eta)$ in the integrand of  \eqref{eq:12} to obtain
\begin{eqnarray}\label{eq:140}
\langle (\Delta v^z)^{2}\rangle &=&\dfrac{q^{2}}{m^{2}a_{f}^{2}}\left(1+\dfrac{3}{2}(\vec{\theta}\cdot\vec{B})\right) \iint d\eta_{1}d\eta_{2}a^{2}(\eta_{2})\left[\Theta^{zx}\partial_{z_1}\partial_{x_2}+\Theta^{zy}\partial_{z_1}\partial_{y_2}\right.\nonumber \\
&+&\left.\Theta^{zz}\partial_{z_1}\partial_{z_2}\right]        {{     \langle\phi(\eta_1, r_1)\phi(\eta_2, r_2)\rangle_{_{M}}          }}   +\dfrac{q^{2}}{m^{2}a_{f}^{2}}\left(1+\dfrac{3}{2}(\vec{\theta}\cdot\vec{B})\right) \iint d\eta_{1}d\eta_{2}a^{2}(\eta_{1})\left[\Theta^{zx}\partial_{x_1}\partial_{z_2}\right.\nonumber\\
&+&\left.\Theta^{zy}\partial_{y_1}\partial_{z_2}+\Theta^{zz}\partial_{z_1}\partial_{z_2}\right]
                                                               {{     \langle\phi(\eta_1, r_1)\phi(\eta_2, r_2)\rangle_{_{M}}         }}\nonumber \\
&+&\dfrac{q^{2}}{m^{2}a_{f}^{2}}\iint d\eta_{1}d\eta_{2}a^{2}(\eta_{1})a^{2}(\eta_{2})\{\Theta^{zx}[\Theta^{zx}\partial_{x_1}\partial_{x_2}+\Theta^{zy}\partial_{x_1}\partial_{y_2}
+\Theta^{zz}\partial_{x_1}\partial_{z_2}]\nonumber\\
&+& \Theta^{zy}[\Theta^{zx}\partial_{y_1}\partial_{x_2}+\Theta^{zy}\partial_{y_1}\partial_{y_2}
+\Theta^{zz}\partial_{y_1}\partial_{z_2}] + \Theta^{zz}[\Theta^{zx}\partial_{z_1}\partial_{x_2}+\Theta^{zy}\partial_{z_1}\partial_{y_2}\nonumber\\
&+&\Theta^{zz}\partial_{z_1}\partial_{z_2}]\}                  {{     \langle\phi(\eta_1, r_1)\phi(\eta_2, r_2)\rangle_{_{M}}           }}.
\end{eqnarray}
Because the integrals above has at least one $a^2(\eta)$ factor in their integrands, they give finite contributions. We also use the fact that $\theta$ is small and the following  Taylor expansion was made:
\begin{equation}
H^{-2}\left(1 - \frac{1}{2}{ \vec{\theta}\cdot\vec{B}}\right) 
\approx 1 + \frac{3}{2}\left(\vec{\theta}\cdot\vec{B}\right) 
+ \frac{9}{2}\left(\vec{\theta}\cdot\vec{B}\right)^2+\cdots ,
\end{equation}
where $H$ is given by Eq. (\ref{eq:HW}) and terms up to second  order were considered. Note that the derivatives above obey the relation,

\begin{equation}\label{eq:15}
\partial_{k_1}\partial_{\ell_2}\langle\phi_{1}\phi_{2}\rangle_{_{M}}=\left\{
\begin{array}{cc}
\textrm{If}\ k=\ell, & \dfrac{1}{2\pi^{2}}[f_{2}(\eta,r)+4\Delta k^{2}f_{3}(\eta,r)]\\
\textrm{If}\ k\neq\ell, & \dfrac{2\Delta k\Delta\ell f_{3}(\eta,r)}{\pi^{2}}
\end{array}\right.,
\end{equation}
%with $k,\ell=x, y, z$; $\Delta k=k_{1}-k_{2}$; $\Delta \ell=\ell_{1}-\ell_{2}$ and %
where $\Delta k=k_{1}-k_{2}$, $\Delta \ell=\ell_{1}-\ell_{2}$, with $k,\ell=x, y, z$, and
\begin{equation}\label{eq:151}
f_{n}(\eta, r)=\dfrac{1}{[c_{0}^{2}(\eta_{1}-\eta_{2})^{2}-r^{2}]^{n}},
\end{equation}
%where $n=2,3$ and $ r^2= (x_{1}-x_{2})^{2}+(y_{1}-y_{2})^{2}+(z_{1}-z_{2})^{2}=\Delta x^{2}+\Delta y^{2}+\Delta z^{2}$ with $\partial_{k_1}\partial_{\ell_2}\langle\phi_{1}\phi_{2}\rangle_{_{M}}=\partial_{\ell_1}\partial_{k_2}\langle\phi_{1}\phi_{2}\rangle_{_{M}}$. %
with $n=2,3$ and $r$ defined by Eq.~\eqref{eq:fp1}.

Although the number of terms is significantly large, from  the generalization given by Eq. \eqref{eq:15}, we note that some terms will be null in the coincidence limit ($r\rightarrow 0$). 
%Moreover, since $(\eta_1-\eta_2)^{4}=(\eta_2-\eta_1)^{4}$, the limits and factors that multiply the first two right-hand integrals of Eq. \eqref{eq:140} are equal, the change between them occurs only on the label. 
%So we need to solve one of these integrals and multiply by two. 
Thus,  applying this limit we have
\begin{eqnarray}\label{eq:16}
\langle (\Delta v^z)^{2}\rangle &=&\dfrac{q^{2}\Theta^{zz}}{\pi^{2}m^{2}a_{f}^{2}c_{0}^{4}}\left( 1 +\dfrac{3}{2}(\vec{\theta}\cdot\vec{B})\right)\int_{0}^{\eta_{f}}d\eta_{2}\int_{-\infty}^{\infty}d\eta_{1}a^{2}(\eta_{1})\dfrac{1}{(\eta_{1}-\eta_{2})^{4}}\nonumber \\
&+& \dfrac{q^{2}}{2\pi^{2}m^{2}a_{f}^{2}c_{0}^{4}}\int_{-\infty}^{\infty}d\eta_{2}a^{2}(\eta_{2})\int_{-\infty}^{\infty}d\eta_{1}a^{2}(\eta_{1})\left[\dfrac{\Theta^{zx}\Theta^{zx}}{(\eta_{1}-\eta_{2})^{4}}\right.\nonumber\\
&+&\left. \dfrac{\Theta^{zy}\Theta^{zy}}{(\eta_{1}-\eta_{2})^{4}}+\dfrac{\Theta^{zz}\Theta^{zz}}{(\eta_{1}-\eta_{2})^{4}} \right].
\end{eqnarray}
Using the scale factor given by Eq.~\eqref{eq:fatesc} into Eq.~\eqref{eq:16}  we first integrates by parts and then uses the residue theorem  \cite{boas2006}, to obtain  the velocity dispersion of the particles in the $z$-direction
\begin{eqnarray}\label{eq:18a}
\langle (\Delta v^z)^{2}\rangle &= &\dfrac{2q^{2}a_{1}^{2}}{3\pi^{4}m^{2}a_{f}^{2}c_{0}^{4}\eta_{0}^{2}}(\theta^{x}B^{x}+\theta^{y}B^{y})\left(1 +\dfrac{3}{2}(\vec{\theta}\cdot\vec{B})\right)\nonumber\\
&\times &\left\{7\zeta(3)+Re\left[\frac{1}{2}\Psi\left(2,\dfrac{\pi+2wi}{2\pi}\right)\right]\right\}\nonumber \\
&+& \dfrac{2q^{2}a_{1}^{4}}{\pi^{4}m^{2}a_{f}^{2}c_{0}^{4}\eta_{0}^{2}}\zeta(3) \left[(\theta^{x}B^{z})^{2}+(\theta^{y}B^{z})^{2}+(\theta^{x}B^{x}+\theta^{y}B^{y})^{2}\right],
\end{eqnarray}
where we have used the  definition $w=\eta_{f}/\eta_{0}$, which represents a relation between the final conformal time $\eta_{f}$ and the constant parameter $\eta_{0}$. Here, $\zeta (x)$ is the zeta function and $\Psi (n, x)$ is the nth-polygamma function.

As we are dealing with $\vec{B}=B^{z}\hat{k}$ only, the above expression simplifies to  
\begin{eqnarray}\label{eq:18b}
\langle (\Delta v^z)^{2}\rangle = \dfrac{2q^{2}a_{1}^{4}}{\pi^{4}m^{2}a_{f}^{2}c_{0}^{4}\eta_{0}^{2}}\zeta(3) \left[(\theta^{x}B^{z})^{2}+(\theta^{y}B^{z})^{2}\right],
\end{eqnarray}
which corresponds to the dispersion parallel to the magnetic field.  Note that, the velocity dispersion in the $z$-direction is constant and different from zero. The dominant noncommutativity contribution is a second order term and when $\theta^i = 0$ or $B^z = 0$ the dispersion is null which is the same result found in Ref. \cite{Bessa2017} for a commutative expanded fluid.  

Now let us apply the same methodology for the perpendicular dispersion. For this we consider $i=x$  in Eq.~\eqref{eq:12}, to obtain\footnote{The case $i=y$ is also perpendicular to the field and it gives similar results.  } 
\begin{eqnarray}\label{eq:19}
\langle (\Delta v^x)^{2}\rangle &=&\dfrac{q^{2}\Theta^{xx}}{\pi^{2}m^{2}a_{f}^{2}c_{0}^{4}}\left( 1 +\dfrac{3}{2}(\vec{\theta}\cdot\vec{B})\right)\int_{0}^{\eta_{f}}d\eta_{2}\int_{-\infty}^{\infty}d\eta_{1}a^{2}(\eta_{1})\dfrac{1}{(\eta_{1}-\eta_{2})^{4}}\nonumber \\
&+& \dfrac{q^{2}}{2\pi^{2}m^{2}a_{f}^{2}c_{0}^{4}}\int_{-\infty}^{\infty}d\eta_{2}a^{2}(\eta_{2})\int_{-\infty}^{\infty}d\eta_{1}a^{2}(\eta_{1})\left[\dfrac{\Theta^{xx}\Theta^{xx}}{(\eta_{1}-\eta_{2})^{4}}\right.\nonumber\\
&+&\left. \dfrac{\Theta^{xy}\Theta^{xy}}{(\eta_{1}-\eta_{2})^{4}}+\dfrac{\Theta^{xz}\Theta^{xz}}{(\eta_{1}-\eta_{2})^{4}} \right].
\end{eqnarray}
Note that we have used the fact that the two first integrals are equal.
Since  $\Theta$ is constant, the integrals of Eq.~\eqref{eq:19} are the same as the ones obtained in Eq.~\eqref{eq:18a}. The difference appears in the $\Theta$ factors, which by the use of Eq.~\eqref{eq:06}, they can be expressed in terms of $\vec{\theta}$ and $\vec{B}$. Thus, using the scale factor \eqref{eq:fatesc}, we get
\begin{eqnarray}\label{eq:20}
\langle (\Delta v^x)^{2}\rangle &=&\dfrac{2q^{2}a_{1}^{2}}{3\pi^{4}m^{2}a_{f}^{2}c_{0}^{4}\eta_{0}^{2}}(\theta^{y}B^{y}+\theta^{z}B^{z})\left(1 +\dfrac{3}{2}(\vec{\theta}\cdot\vec{B})\right)\nonumber\\
&\times &\left\{7\zeta(3)+Re\left[\frac{1}{2}\Psi\left(2,\dfrac{\pi+2wi}{2\pi}\right)\right]\right\}\nonumber \\
&+& \dfrac{2q^{2}a_{1}^{4}}{\pi^{4}m^{2}a_{f}^{2}c_{0}^{4}\eta_{0}^{2}}\zeta(3) \left[(\theta^{y}B^{y}+\theta^{z}B^{z})^{2}+(\theta^{y}B^{x})^{2}+(\theta^{z}B^{x})^{2}\right],
\end{eqnarray}
which is the expression for the proper velocity dispersion in $x$ direction  and $w = \eta_f/\eta_0$. {The fact that the  magnetic field $\vec{B}=B^{z}\hat{k}$ results in }

\begin{figure}[htbp]
\centering
\includegraphics[scale=0.5]{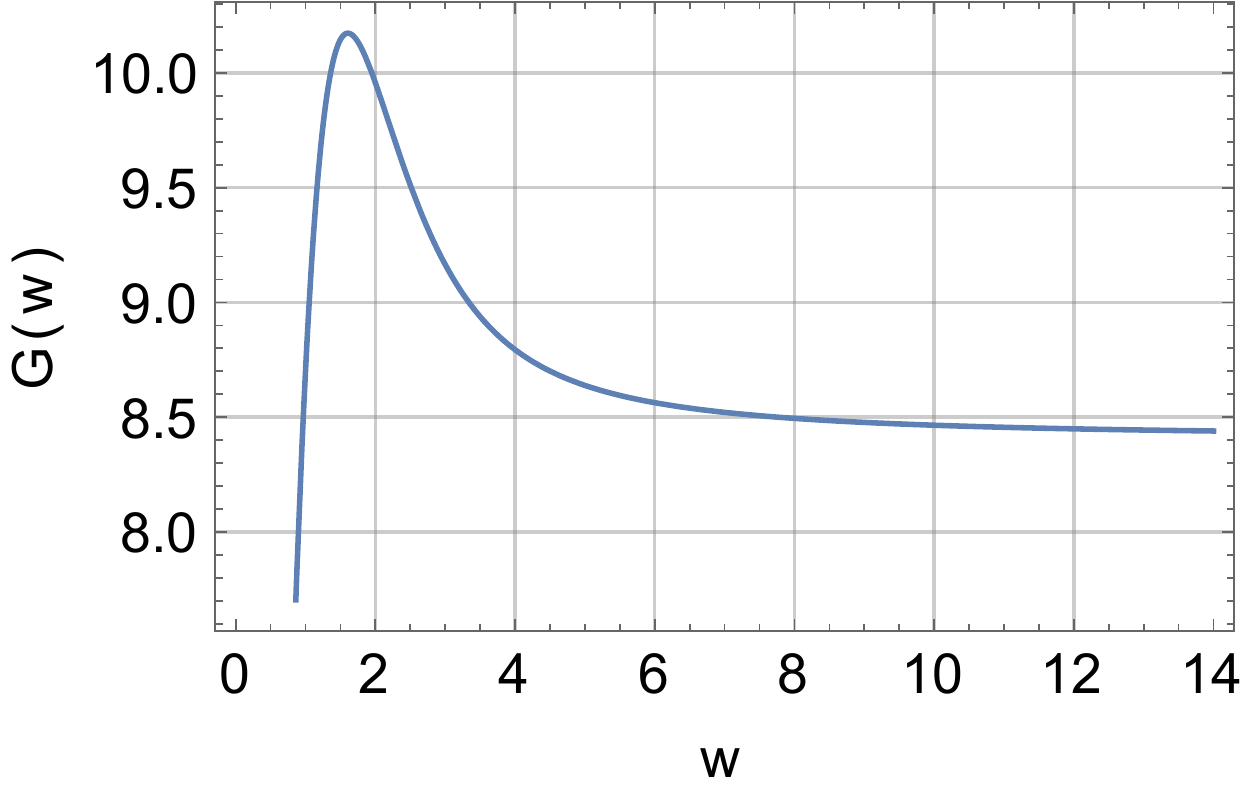} 
\caption{The contribution produced by noncommutativity in the equation \eqref{eq:21b} as a function of the $w$. Here, $w=\eta_{f}/\eta_{0} > 0$ and $G(w)\equiv \langle (\Delta v^x)^{2}\rangle\times[ 3\pi^{4}m^{2}a_{f}^{2}c_{0}^{4}\eta_{0}^{2}/2q^{2}a_{1}^{2}(\theta^{z}B^{z})]$.}\label{figgcaliv}
\end{figure}

%which is the general expression to the proper velocity dispersion of the particles in $x$ direction, which with the particularization for the magnetic field $\vec{B}=B_{z}\hat{k}$ becomes %
\begin{eqnarray}\label{eq:21a}
\langle (\Delta v^x)^{2}\rangle &=&\dfrac{2q^{2}a_{1}^{2}}{3\pi^{4}m^{2}a_{f}^{2}c_{0}^{4}\eta_{0}^{2}}(\theta^{z}B^{z})\left( 1 +\dfrac{3}{2}(\theta^{z}B^{z})\right)\nonumber\\
&\times &\left\{7\zeta(3)+Re\left[\frac{1}{2}\Psi\left(2,\dfrac{\pi+2wi}{2\pi}\right)\right]\right\}\nonumber \\
&+& \dfrac{2q^{2}a_{1}^{4}}{\pi^{4}m^{2}a_{f}^{2}c_{0}^{4}\eta_{0}^{2}}\zeta(3)(\theta^{z}B^{z})^{2},
\end{eqnarray}
or considering up to the first order terms in $\theta$,

\begin{eqnarray}\label{eq:21b} 
\langle (\Delta v^x)^{2}\rangle &\approx &\dfrac{2q^{2}a_{1}^{2}}{3\pi^{4}m^{2}a_{f}^{2}c_{0}^{4}\eta_{0}^{2}}(\theta^{z}B^{z})\left\{7\zeta(3)+Re\left[\frac{1}{2}\Psi\left(2,\dfrac{\pi+2wi}{2\pi}\right)\right]\right\}.
\end{eqnarray}
Note that now the velocity dispersion in the $x$ direction has a time dependence. This dependence in time did not appear in the $z$-direction. 
It is  given in terms of the dimensionless parameter $w=\eta_f/\eta_0$.
Fig.~\ref{figgcaliv} shows curves for Eq.~\eqref{eq:21b} as function of the $w$ parameter.

\begin{figure}[htbp]
\centering
\includegraphics[scale=0.5]{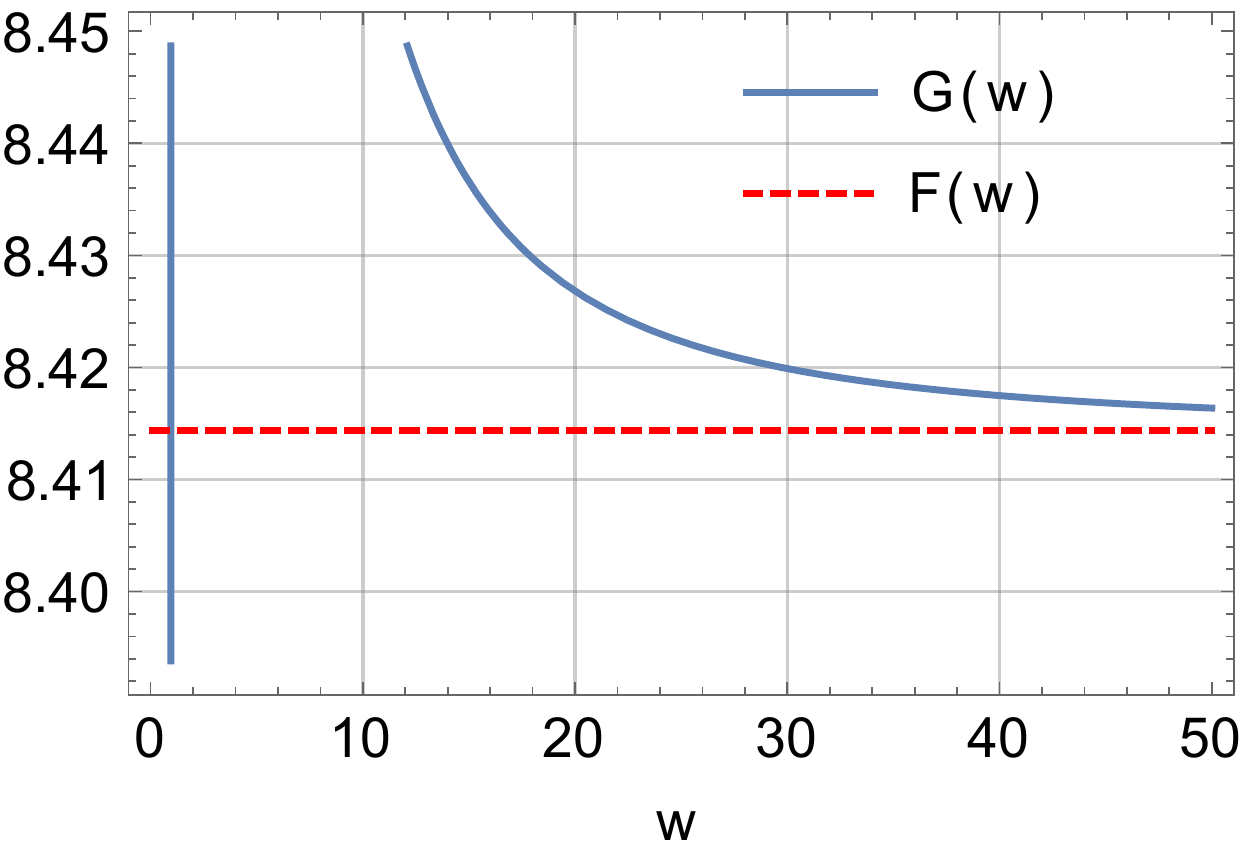}
\caption{The contribution produced by noncommutativity in the equation \eqref{eq:21b} (represented by $G(w)$) and \eqref{eq:22} (represented by $F(w)$) as a function of $w$. Here, $w=\eta_{f}/\eta_{0} > 0$ and  $G(w)\equiv \langle (\Delta v^x)^{2}\rangle\times[ 3\pi^{4}m^{2}a_{f}^{2}c_{0}^{4}\eta_{0}^{2}/2q^{2}a_{1}^{2}(\theta^{z}B^{z})]$ and  $F(w)\approx\langle (\Delta v^x)^{2}\rangle_{Asymptotic}\times[ 3\pi^{4}m^{2}a_{f}^{2}c_{0}^{4}\eta_{0}^{2}/2q^{2}a_{1}^{2}(\theta^{z}B^{z})]$ .}\label{soluanageral}
\end{figure}

From Fig.~\ref{figgcaliv} we note that for large values of $w$, corresponding to large values of time $\eta_{f}$,  the noncommutativity effect approaches a constant value. Moreover, the effect has a fast behavior %("abrupt variation")% 
to short values of $w$, that is, small time separation between $\eta_{f}$ and $\eta_{0}$. However, the most important result here is to find an asymptotic solution for $\langle (\Delta v^x)^{2}\rangle$. Thus for large $w$ we find % Thus, in principle, such effect would be easier to verify in situations where $w$ tends to large values because it tends to a constant value. We must remember that the parameter $\eta_{0}$ is a time constant whose meaning is the fluid expansion rate and no the initial time.

%We can find an approximate solution which describes the asymptotic behavior of $\langle (\Delta v^x)^{2}\rangle$ and an approximate constant value of $G(w)$ as shown in Fig.~\ref{figgcaliv}. Thus, the regime of large $w$ reads %from eq. \eqref{eq:17a} we obtain  
\begin{eqnarray}\label{eq:22}
\langle (\Delta v^x)^{2}\rangle_{Asymptotic} \approx  \dfrac{14q^{2}a_{1}^{2}}{3\pi^{4}m^{2}a_{f}^{2}c_{0}^{4}\eta_{0}^{2}}(\theta^{z}B^{z})\zeta(3).
\end{eqnarray}
In the Fig.~\ref{soluanageral} we show the plots of the asymptotic result \eqref{eq:22} (dashed line), together with \eqref{eq:21b}  (solid line). 

%From Fig.~\ref{soluanageral} we note that the equation \eqref{eq:22} present a satisfactory description for the velocity dispersion behavior  of the particles when $w$ tends to large values. %, that is, the equation \eqref{eq:22} is used to describe only the asymptotic behavior of $\langle (\Delta v^x)^{2}\rangle$, since for short values of $w$ we have a curve that deviates from this regime.

In summary,  Eqs. \eqref{eq:18a} and \eqref{eq:20}  are the general expressions for the velocity dispersion of the massive scalar particles that form the noncommutative fluid in the $z$ and $x$ directions,  respectively.  When we assume that the magnetic field is nonzero only in the $z$-direction we obtain Eqs. \eqref{eq:18b}, (\ref{eq:21a}). So, these equations show a non-null velocity dispersion. This result disagrees with the one found in Ref. \cite{Bessa2017}, where the stochastic motion of scalar particles with mass m for a commutative fluid was studied. In this case, the fluid considered was a BEC and in a similar way to what was done in Sect. \ref{sec_metric} of the present paper,  linear perturbations in the fluid parameters $(\rho, \phi, S)$ were applied, and a metric similar to the one found in Eq.  (\ref{eq:02}) was obtained with the same scale factor present in Eq. (\ref{eq:fatesc}). {Thus, due to the null result found in Ref. \cite{Bessa2017}, the non-null result found in the present paper, suggests a relevant contribution coming from the noncommutativity in the stochastic motion of the particles.}% induced by a fluctuating quantized scalar field in this non-static spacetime, which is a non-trivial quantum effect,}% we suggest that the results found here could give a piece of evidence for the noncommutativity of space. 

%\large{\textbf{Thus, the non-zero result found in this section could be a piece of evidence for the noncommutativity of space.    }} 

\subsection{Bound particle in an expanding noncommutative fluid}

Similarly to the previous case, we will now analyze the velocity dispersion in two distinct directions. First, let us verify the dispersion in $z$ direction. From equation \eqref{eq:23} we obtain
\begin{eqnarray}\label{eq:24}
\langle (\Delta v^z)^{2}\rangle &=&\dfrac{q^{2}a^{2}_{f}}{2\pi^{2}m^{2}}(1+3\vec{\theta}\cdot\vec{B})\iint d\eta_{1}d\eta_{2}a^{-2}(\eta_1)a^{-2}(\eta_2)[f_{2}(\eta,z)+4\Delta z^{2}f_{3}(\eta,z)]\nonumber \\
&+&\dfrac{2q^{2}a^{2}_{f}}{\pi^{2}m^{2}}\iint d\eta_{1}d\eta_{2}a^{-2}(\eta_1)\bigg\{2\Theta^{zx}\Delta x\Delta zf_{3}(\eta,z)+2\Theta^{zy}\Delta y\Delta zf_{3}(\eta,z)\nonumber\\
&+&\dfrac{1}{2}\Theta^{zz}[f_{2}(\eta,z)+4\Delta z^{2}f_{3}(\eta,z)]\bigg\},
\end{eqnarray}
where we have used the definition of $f_{n}(\eta,z)$ given by  Eq. \eqref{eq:151}, and the fact that the second and third integrals in the right hand side of Eq. (\ref{eq:23})  are equal and  a Taylor expansion in the noncommutative parameter ($\theta$), up to first order, was made. Furthermore, in the fourth term in the right hand side of Eq. (\ref{eq:23}) the Minkowski vacuum divergent term must be subtracted during the renormalization procedure. 

Finally, applying the coincidence limit in Eq. (\ref{eq:24}) we find, 

\begin{eqnarray}\label{eq:24a}
\langle (\Delta v^z)^{2}\rangle &=&\dfrac{q^{2}a^{2}_{f}}{2\pi^{2}m^{2}c_{0}^{4}}(1+3\vec{\theta}\cdot\vec{B})\int_{-\infty}^{+\infty} a^{-2}(\eta_2) \int_{-\infty}^{+\infty} d\eta_{1}a^{-2}(\eta_1)\dfrac{1}{(\eta_{1}-\eta_{2})^{4}}\nonumber \\
&+&\dfrac{q^{2}a^{2}_{f}\Theta^{zz}}{\pi^{2}m^{2}c_{0}^{4}}\int_{0}^{\eta_{f}} d\eta_{2} \int_{-\infty}^{+\infty} d\eta_{1}a^{-2}(\eta_1)\dfrac{1}{(\eta_{1}-\eta_{2})^{4}}.
\end{eqnarray}
Using the scale factor given by Eq.~\eqref{eq:fatesc} and the residue integration method  in Eq. \eqref{eq:24a} we obtain
\begin{eqnarray}\label{eq:26a}
\langle (\Delta v^z)^{2}\rangle &=&\dfrac{2q^{2}a^{2}_{f}\sinh^{4}(g)}{\pi^{4}m^{2}\eta_{0}^{2}a_{1}^{4}c_{0}^{4}}\left[\zeta(3)-\dfrac{\pi^{4}}{90}\right](1+3\vec{\theta}\cdot\vec{B}) + \dfrac{q^{2}a^{2}_{f}\sinh^{2}(g)}{3\pi^{4}m^{2}\eta_{0}^{2}a_{1}^{2}c_{0}^{4}}(\theta^{x}B^{x}+\theta^{y}B^{y})\nonumber\\
&\times &Re\left[\Psi\left(2,\dfrac{\pi+2gi}{2\pi}\right)-\Psi\left(2,\dfrac{\pi+2(g+w)i}{2\pi}\right)\right],
\end{eqnarray}
where $w=\eta_{f}/\eta_{0}$ and
\begin{eqnarray}
g=\dfrac{1}{2}\ln\left(\dfrac{\alpha^{2}+1}{\alpha^{2}-1}\right) = \dfrac{1}{2}\ln\left(\dfrac{c_{0}}{c_{sf}}\right)\label{eq:25e},
\end{eqnarray}
with $\alpha^{2} = a_{0}^{2}/a_{1}^{2} > 1$. The parameter $c_{sf}=c_{s}(\eta=\eta_f)$ represents the final sound velocity in the fluid at a final time $\eta_f$ and $c_{0}$ the initial sound velocity. %Since $a(\eta) = c_{0}/c_{s}(t)$, it is important to note that $c_{sf}<c_{0}$ is a necessary condition to \large{\textbf{the}} expansion occur, that is, for $a(\eta)$ assuming increasing values.

The equation \eqref{eq:26a} is the general expression to the proper velocity dispersion of the particles in $z$ direction.
%As before, by choosing the magnetic field in the $z$ direction, we get 
Applying the same procedure done in the previous section, let us choose the magnetic field in the $z$ direction, so we get
\begin{eqnarray}\label{eq:26b}
\langle (\Delta v^z)^{2}\rangle =\dfrac{2q^{2}a_{f}^{2}\sinh^{4}(g)}{m^{2}\pi^{4}a_{1}^{4}c_{0}^{4}\eta_{0}^{2}}\left[\zeta(3)-\dfrac{\pi^{4}}{90}\right](1+3\theta^{z}B^{z}).
\end{eqnarray}
Note that, for $\theta^z = 0$ or $B^z = 0$ we recover the result found in Ref. \cite{Bessa2017} in the absence of boundaries. Thus, in the present paper, there is an additional first order contribution due the noncommutativity of space.  

%The  \large{\textbf{result found above}} is similar to that \large{\textbf{one}} found  \large{\textbf{previously}} in the literature for the velocities dispersion of the particles in the absence of boundary \cite{Bessa2017}. \large{\textbf{However, now there is an additional term that comes from the non-commutative correction.}}%But, now there is the presence of an additional term coming from the non-commutative correction.

Now, for completeness, let us investigate the velocity dispersion in a perpendicular direction, by making $i=x$ in the equation \eqref{eq:23}, we obtain
\begin{eqnarray}
\langle (\Delta v^x)^{2}\rangle &=&\dfrac{q^{2}a^{2}_{f}}{2\pi^{2}m^{2}}(1+3\vec{\theta}\cdot\vec{B})\iint d\eta_{1}d\eta_{2}a^{-2}(\eta_1)a^{-2}(\eta_2)[f_{2}(\eta,x)+4\Delta x^{2}f_{3}(\eta,x)]\nonumber \\
&+&\dfrac{2q^{2}a^{2}_{f}}{\pi^{2}m^{2}}\iint d\eta_{1}d\eta_{2}a^{-2}(\eta_1)\bigg\{\dfrac{1}{2}\Theta^{xx}[f_{2}(\eta,x)+4\Delta x^{2}f_{3}(\eta,x)]\nonumber\\
&+&2\Theta^{xy}\Delta x\Delta yf_{3}(\eta,x)+2\Theta^{xz}\Delta x\Delta zf_{3}(\eta,x)\bigg\}.
\end{eqnarray}
Since the  factors $\Theta^{ij}$ are constants  the integrals in the $x$ direction are equal to those obtained to the $z$ direction. Therefore in the coincidence limit
\begin{eqnarray}\label{eq:27a}
\langle (\Delta v^x)^{2}\rangle &=&\dfrac{2q^{2}a^{2}_{f}\sinh^{4}(g)}{\pi^{4}m^{2}\eta_{0}^{2}a_{1}^{4}c_{0}^{4}}\left[\zeta(3)-\dfrac{\pi^{4}}{90}\right](1+3\vec{\theta}\cdot\vec{B})+\dfrac{q^{2}a^{2}_{f}\sinh^{2}(g)}{3\pi^{4}m^{2}\eta_{0}^{2}a_{1}^{2}c_{0}^{4}}(\theta^{y}B^{y}+\theta^{z}B^{z})\nonumber\\
&\times &Re\left[\Psi\left(2,\dfrac{\pi+2gi}{2\pi}\right)-\Psi\left(2,\dfrac{\pi+2(g+w)i}{2\pi}\right)\right],
\end{eqnarray}
and once we choose the magnetic field in the $z$ direction
\begin{eqnarray}\label{eq:27b}
\langle (\Delta v^x)^{2}\rangle &=&\dfrac{2q^{2}a^{2}_{f}\sinh^{4}(g)}{\pi^{4}m^{2}\eta_{0}^{2}a_{1}^{4}c_{0}^{4}}\left[\zeta(3)-\dfrac{\pi^{4}}{90}\right](1+3\theta^{z}B^{z})+\dfrac{q^{2}a^{2}_{f}\sinh^{2}(g)}{3\pi^{4}m^{2}\eta_{0}^{2}a_{1}^{2}c_{0}^{4}}(\theta^{z}B^{z})\nonumber\\
&\times &Re\left[\Psi\left(2,\dfrac{\pi+2gi}{2\pi}\right)-\Psi\left(2,\dfrac{\pi+2(g+w)i}{2\pi}\right)\right].
\end{eqnarray}
In this particular case we have a time dependence given in terms of the parameter $w$. Now, let us plot the effect of the non-commutativity. For this purpose, we define a new function $T(w)$%a temporal dependence through the parameter $w$. As before we will draw a graph with the purpose to verify the effects from the noncommutative terms. For this, we define a function 
\begin{eqnarray}\label{eq:28a}
T(w)&=& \dfrac{6\sinh^{2}(g)}{a_{1}^{2}}\left[\zeta(3)-\dfrac{\pi^{4}}{90}\right] \nonumber \\
&+&\dfrac{1}{3}Re\left[\Psi\left(2,\dfrac{\pi+2gi}{2\pi}\right)-\Psi\left(2,\dfrac{\pi+2(g+w)i}{2\pi}\right)\right],
\end{eqnarray}
this is the term proportional to $\theta^z B^z$ and it can be written as
\begin{eqnarray}\label{eq:28b}
T= \dfrac{\pi^{4}m^{2}a_{1}^{2}c_{0}^{4}\eta_{0}^{2}}{q^{2}a_{f}^{2}\sinh^{2}(g)(\theta^{z}B^{z})}\left\{\langle (\Delta v^x)^{2}\rangle - \dfrac{2q^{2}a_{f}^{2}\sinh^{4}(g)}{m^{2}\pi^{4}a_{1}^{4}c_{0}^{4}\eta_{0}^{2}}\zeta(3)\right\}.
\end{eqnarray}
As we know from \eqref{eq:25e}, the $g$ factor is expressed in terms of the fluid parameter and similarly we can show that
\begin{equation}
\dfrac{\sinh^{2}(g)}{a_{1}^{2}} = \dfrac{1}{2}\left(1-\dfrac{c_{sf}}{c_{0}}\right).
\end{equation}

\begin{figure}[hbtp]
\centering
\includegraphics[scale=0.5]{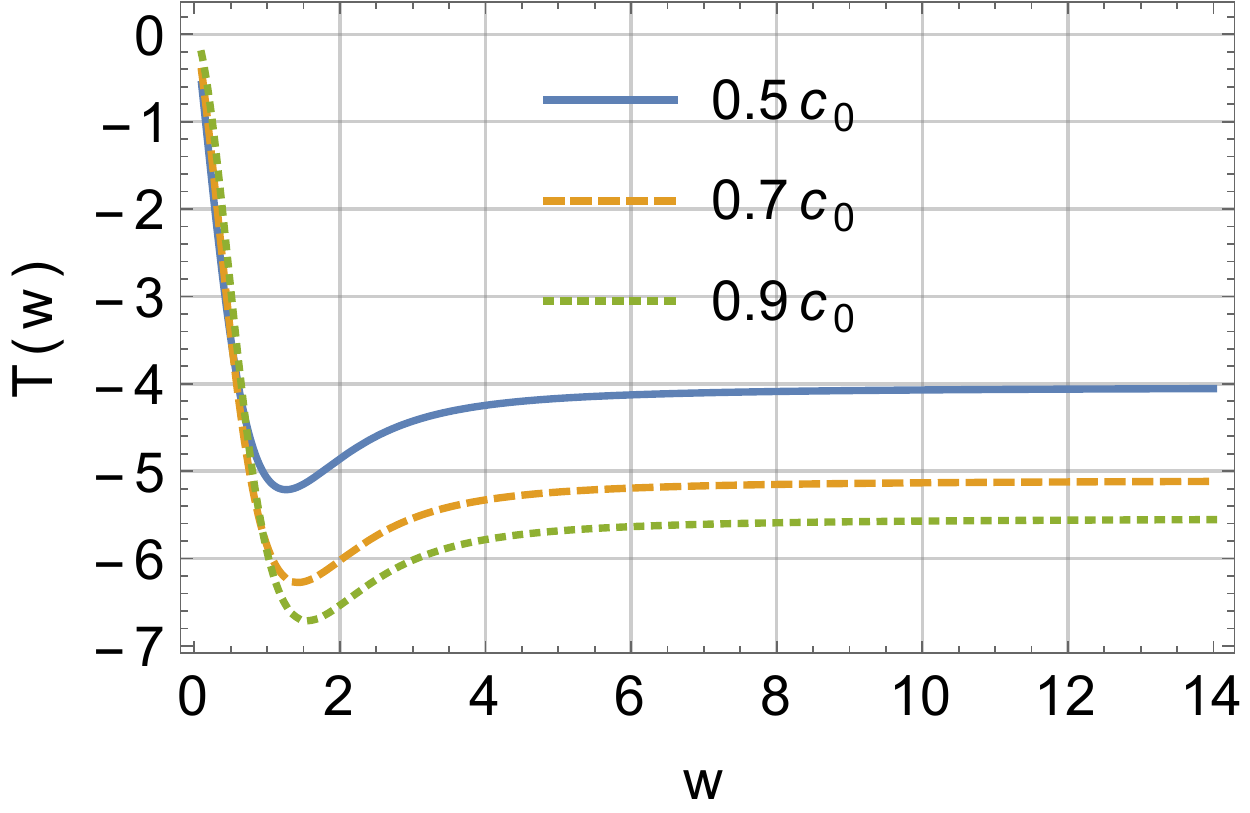}
\caption{The contribution of the noncommutativity in equation \eqref{eq:28a}. The three lines represents the curves for distinct values that the final velocity $c_{sf}$ take in relation to the initial fluid velocity $c_{0}$.}\label{GrafCasLig01}
\end{figure}

\begin{figure}[hbtp]
\centering
\includegraphics[scale=0.35]{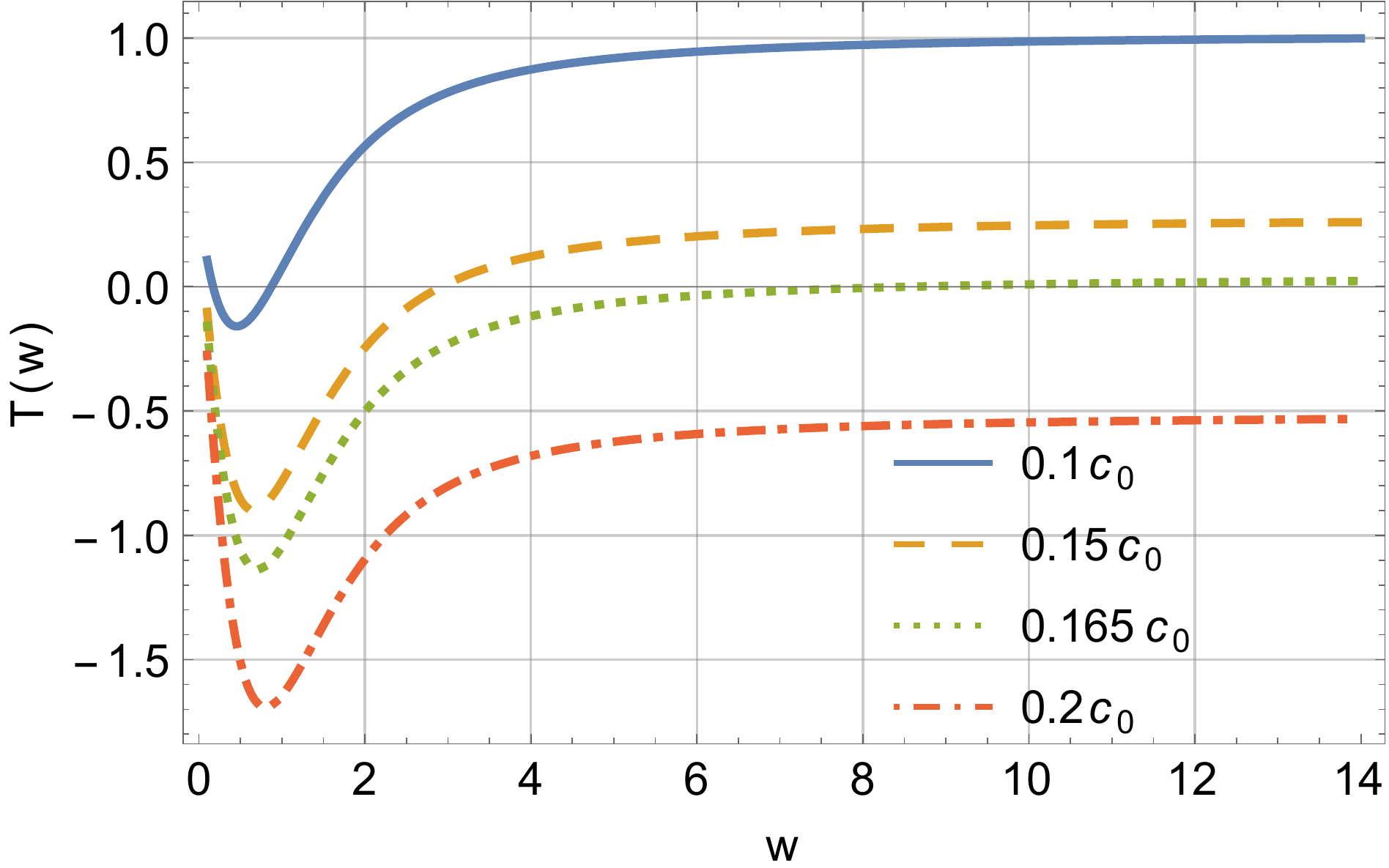}
\caption{The contribution of the noncommutativity in equation \eqref{eq:28a} for small values of $c_{sf}$.}\label{GrafCasLig02}
\end{figure}

 Figure \ref{GrafCasLig01} shows the plot of Eq. (\ref{eq:28a}) for different values of $c_{sf}$. We observe that the first order noncommutativity contribution  on the velocity dispersion in the $x$ direction (perpendicular to the magnetic field) is negative. Note that, $T(w)$ decreases for small values of $w$, but when $w$ assumes large values we note that $T(w)$ tends to a constant. In addition, when the sound velocity in the fluid $c_{sf}$ assumes large values, the noncommutative corrections are more negative. It also shows the influence of noncommutativity on the velocity dispersion of the particles when the sound velocity in fluid is relatively large, however when the sound velocity in fluid $c_{sf}$ takes on small values, as can be seen  in Fig. \ref{GrafCasLig02} in a region of large $w$ and small $c_{sf}$, the noncomutativity contributions for velocity dispersion can be positive.%but something interesting happens when the sound velocity in fluid $c_{sf}$ takes small values. As shown in Fig.~\ref{GrafCasLig02} in a region of large $w$ values for a specific range of $c_{sf}$ values, the noncommutativity contributions for velocity dispersion are positive. 

 As it was noted in Ref. \cite{Bessa2017} for  commutative spaces with bounded particles the velocity dispersion of the particles is constant and is the same in all directions, that is,  in the absence of boundary it is isotropic. However, as shown in Eqs. \eqref{eq:26b} and \eqref{eq:27b},  for noncommutative spaces we obtain different velocity dispersion for different directions and the isotropy, at least in first order, is broken. It is also important to note that the negative contributions that appears in Figs. \ref{GrafCasLig01} and \ref{GrafCasLig02} are up to the first order $\theta$-corrections. The dominant contribution is really the zeroth order term present in Eq. (\ref{eq:27b}) which is positive.

\section{Metric Fluctuations}\label{Sec:fluctuations}

{{

 In this section, we wish to discuss briefly the validity of the methodology adopted in previous sections where the effects of the metric fluctuations were not considered. We will see that these effects can be, for the cases treated in this paper, neglected. So, let us remember that at the end of Sect. \ref{random FRW}, we have said that these fluctuations are in general viewed as linearized perturbations upon a metric. In some analog models for quantum gravity effects, the metric fluctuations could induce sound cone fluctuations. In general, a diagonal metric is taken into account, and white or colored noise is implemented in the equation of motion of the particles \cite{kms10}. Thus, we could write Eq. (\ref{eq:01}) under the perspective of Eq. (\ref{Eq:perturbation}) in the following diagonal form,  
\begin{equation}\label{linearized1}
ds^2 = H^2(1+\gamma)\left[ -c_0^2dt^2 + a^2(dx^2 + dy^2 + dz^2) \right],
\end{equation}
where, in general, $\gamma = \gamma(t, r)$ and $\gamma \ll 1$. In what follows, let us incorporate these fluctuations on the geodesic equation. Thus, the right hand side (RHS) of Eq. (\ref{eq:fcfq}) becomes

\begin{equation}\label{Eq:geoflu}
\frac{1}{m}f^i \cong \frac{du^i}{dt} + 2\frac{\dot{a}}{a}u^i + \dot{\gamma}u^i + (\partial_j\gamma)u^iu^j,
\end{equation}
where we have considered up to the first order terms in $\gamma$. %Note that the  is Eq. (\ref{eq:fcfq}). 

%We have considered that the particles could be bounded by some external force proportional to $\dot{a}/a$ or, they could be free with no external force. 
Since, from Eq. (\ref{eq:14}), $a \geq 1$, and in many analog models \cite{kms10,akms12, bds12, dms13, adsbm14}, the tensor $\gamma_{\mu\nu}$ has only the $\gamma_{00}$ component, which  is regarded as white noise with spatial dependence ($\gamma=\gamma(r)$) only.  As we are working in a non-relativistic limit, $u^i \ll 1$, we can see that, from Eq. (\ref{Eq:geoflu}),  the dominant contribution comes from the two first terms. Then, the RHS of Eq. (\ref{eq:fcfq}) describes the motion of the particles properly. 

Now we wish to see the consistency in considering the fluctuations only in the scalar field $\phi$ and neglecting the metric fluctuations, as was done in  Sect. \ref{random FRW}. In general, these fluctuations has the following features: $\langle\gamma(t, r)\rangle = 0$ and $\langle\gamma(t_1, r_1)\gamma(t_2, r_2)\rangle \neq 0$, which is similar to the fluctuations in $\phi$. 

The $3$-force acting on the particles is given by Eq. (\ref{eq:04}). Note that the metric tensor appears only in the two first terms in the RHS of this equation. So, instead of taking into account metric (\ref{eq:01}), let us consider the linearized metric (\ref{linearized1}). Thus, the $3$-force, up to the first order in $\gamma$ is,

\begin{equation}\label{fgamma}
f^i \cong qH^{-2}a^{-2}\left(1  - \frac{1}{2}\vec{\theta}.\vec{B} - \gamma + \frac{\gamma}{2}\vec{\theta}.\vec{B}\right)\nabla_i\phi + \Theta^{ij}\nabla_j\phi.
\end{equation}
Note that the 4th term  inside parentheses can be neglected since it is proportional to $\theta\gamma$ with both very small.  

%In what follows, we remember that to see the consistency with Ref. \cite{Bessa2017}, we considered in Sect. \ref{Sec.tanh} a null magnetic field. So to make the calculations simpler, let us consider that the magnetic field is off ($B = 0$). Consequently $H = 1$ and Eq. (\ref{fgamma}) reads,

In Sect. \ref{Sec.tanh}, to see the consistency with Ref. \cite{Bessa2017},  a null magnetic field was considered. So to make the calculations simpler, let us consider again that the magnetic field is off ($\vec{B} = 0$). Consequently $H = 1$ and Eq. (\ref{fgamma}) reads,

\begin{equation}
f^i = qa^{-2}(1 - \gamma )\nabla_i\phi.
\end{equation}

The two-point function which must appear in the integrand when the velocity dispersion is being evaluated is now,

%\begin{eqnarray}\label{two-point-gamma}
%\langle f^i(t_1,r_1)f^i(t_2, r_2)\rangle = &&q^2a^{-2}(t_1)a^{-2}(t_2)\left\{\partial_{i_1}\partial_{j_2}\langle\phi(t_1, r_1)\phi(t_2, r_2)\rangle \right. \\ \nonumber &&\left. - \partial_{i_1}\partial_{j_2}\left[\langle\phi(t_1, r_1)\phi(t_2, r_2)\rangle\right]\langle\gamma(t_1, r_1)\gamma(t_2, r_2)\rangle\right\}, 
%\end{eqnarray}
\begin{eqnarray}\label{two-point-gamma}
\langle f^i(t_1,r_1)f^i(t_2, r_2)\rangle \propto \partial_{i_1}\partial_{j_2}\langle\phi(t_1, r_1)\phi(t_2, r_2)\rangle  - \partial_{i_1}\partial_{j_2}\left[\langle\phi(t_1, r_1)\phi(t_2, r_2)\rangle\right]\langle\gamma(t_1, r_1)\gamma(t_2, r_2)\rangle, 
\end{eqnarray}
where we used the fact that $\langle\phi(t_1, r_1)\phi(t_2, r_2)\gamma(t_1, r_1)\gamma(t_2, r_2)\rangle \propto \langle\phi(t_1, r_1)\phi(t_2, r_2)\rangle\langle\gamma(t_1, r_1)\gamma(t_2, r_2)\rangle$ with $\langle\gamma\phi\rangle = 0$. Note that since the second term in the RHS of Eq. (\ref{two-point-gamma}) is a product of fluctuations it is much smaller than the first one.  

In what follows in the rest of  Sect. \ref{Sec.tanh},  when $\vec{B} \neq 0$, it is also possible to neglect this contribution since, in our model, $\vec{B}$ and $\vec{\theta}$ are  non-fluctuating constants. So the two-point functions that appear in the integrands of Sect. \ref{Sec.tanh} is always dominant against the quadratic term of Eq. (\ref{two-point-gamma}) that would appear in these integrals if the metric fluctuations were considered. 

}}

\section{Conclusions}\label{concl}

In this paper, we studied the stochastic motion of a classical scalar particle coupled to a quantized massless scalar field in an expanding noncommutative background. We have shown that this expansion is analogous to a Friedmann-Robertson-Walker  (FRW) geometry. 
To perform this analogy, we considered a decomposition and a linear expansion of the scalar field that is a solution consistent with a Lagrangian describing the noncommutative Abelian Higgs model in a flat spacetime.  In order to simplify our expressions, the model admitted that the electric field present in Lagrangian \eqref{eq00} and the flux velocity were null, and that the noncommutativity was activated when a magnetic field is turned on in a given direction.   Thus, the study for the stochastic motion of a scalar particle was implemented for this expanding noncommutative background. 
It was found  noncommutative  correction for the free and bound particles. Whereas the former were defined as particles that follow geodesics, and the latter as particles that were  under the influence of a classical external force that cancels locally the effects of the expansion. 

 In a recent {paper} \cite{Bessa2017} a commutative Bose-Einstein condensate (BEC)  was taken into account to study the same type of motion described above. In this situation, it was found a non-null velocity dispersion associated with the bound particles, meaning that these particles undergo stochastic motion due to quantum fluctuations.  In the present paper, when bound particles were considered, we have found the same result present in Ref. \cite{Bessa2017} added by a factor  proportional to the noncommutative parameter $\theta$. 
In this case, the noncommutativity contributes as a first-order correction. 
Such correction could be  negative for short times and positive in the long time regime. However, the dominant contribution is the zeroth term in $\theta$, which is positive. 
%These results are illustrated in Figs.~\ref{GrafCasLig01} and \ref{GrafCasLig02}, where it is shown that the dispersion depends on the final sound velocity of the fluid ($c_{sf}$).

In the same context, for a commutative fluid, when no boundary is present, it was found in Ref. \cite{Bessa2017} that the velocity dispersion associated with the free particles was zero.  This means that free particles has no stochastic motion due to the quantized field in the expanding commutative background. However, in the present paper, when the noncommutativity of space is taken into account, a nonzero velocity dispersion was found for the free particles. For the dispersion perpendicular to the magnetic field, it was found a term proportional to $\theta$ and for the dispersion parallel to the magnetic field a term proportional to $\theta^2$ was found.  This result could be interpreted as a direct consequence of the noncommutativity of the space. 

{The stochastic motion of the particles shown in this paper is a subtle, non-trivial quantum effect and, although the mechanism presented here to observe this manifestation is limited, the main interest of our paper is theoretical once our results show a relevant contribution coming from the noncommutativity of space.  }%induced by a fluctuating quantized scalar field in this non-static spacetime, which is a non-trivial quantum effect,}%This fact is illustrated in Fig.~\ref{figgcaliv}. 
%So a possible interpretation from our results is that the commutative BEC is an approximation for a more general noncommutative fluid, like the one proposed in Ref.~\cite{Anacleto2012}. %and Eq. (\ref{eq00}) from the present paper. 

%Considering the great achievement accomplished by the authors of Ref.~\cite{ekjsc18},  a BEC that mimics an expanding universe in the lab is a reality today. Thus, if the proposal discussed in this paper could be implemented in a similar table-top experiment and a non-null velocity dispersion for the free atoms that compound the BEC could be found,  this result  could be a clue of evidence of noncommutativity of space.

\acknowledgments

We would like to thank CNPq, CAPES and CNPq/PRONEX/FAPESQ-PB (Grant nos. 165/2018 and 015/2019),  for partial financial support. MAA, FAB and EP acknowledge support from CNPq (Grant nos. 306962/2018-7 and  433980/2018-4, 312104/2018-9, 304852/2017-1). The authors would like to thank J. P. Spinelly and F. G. Costa for helpful comments.

%%%%%%%%%%%%%%%%%%%%%%%%%%%%%%%%%

\end{document}